\documentclass[12pt]{article}
\usepackage{amssymb}
\usepackage{amsmath}
\usepackage{hyperref}
\begin{document}
\title{\bf Black Brane Solution in Rastall AdS Massive Gravity and Viscosity Bound}
\author{Mehdi Sadeghi$$\thanks{ Email: Mehdi.sadeghi@abru.ac.ir}  \hspace{2mm} \\
     % Shahrokh Parvizi$^{2}$\thanks{Corresponding author: Email:Parvizi@modares.ac.ir}\hspace{2mm}\\
{\small {\em  $$Department of Sciences, University of Ayatollah Ozma Boroujerdi, Boroujerd, Lorestan, Iran}}\\
		%{\small {\em $^2$Department of Physics, School of Sciences,}}\\
       % {\small {\em Tarbiat Modares University, P.O.Box 14155-4838, Tehran, Iran}}\\
       }
\date{\today}
\maketitle

\abstract{In this paper, we introduced the black brane solution in Rastall theory and in the context of massive gravity. The ratio of shear viscosity to entropy density is calculated for this solution. Our result shows that the KSS bound violates for this theory.}\\

%\noindent PACS numbers: 11.10.Jj, 11.10.Wx, 11.15.Pg, 11.25.Tq\\
%\pacs{11.10.Jj, 11.10.Wx, 11.15.Pg, 11.25.Tq}


%--------------------------------------------------------------------------
\section{Introduction} \label{intro}

\indent  General Relativity (GR), formulated by Albert Einstein in 1915, is a theory for massless and spin-2 particles \cite{E1914}--\cite{ESPA1914}. GR is a Gauge Theory and its gauge is diffeomorphism invariance, therefore it should be quantizable but it is not so because GR is non-renormalizable .This problem is solved by adding higher order curvature terms to the Einstein-Hilbert action at 1-loop \cite{KS,I.L.}.\\
 The second reason to modify GR is Dark matter (DM) and Dark energy (DE). The observation shows that $4\% $ of the universe is known and the other $96\%$ of the unknown universe includes DE and DM. Therfore, GR should be modified to describe DM and DE.\\
Rastall theory is one of these modified gravities formulated by P. Rastall.
GR has an assumption that the covariant divergence of the energy-momentum vanishes but the assumption is  modified in the Rastall theory like $T^{\mu}_{\mu ;\nu}=\lambda R_{,\nu}$  where $\lambda$ is the Rastall parameter\cite{Rastall1972}.\\
Field equations reduce to the Einstein equations when $\lambda=0$. These equations are equivalent to the Einstein equations in empty space-time, but differ from them in the presence of matter. The geometry and matter fields are coupled to each other in a non-minimal way in Rastall theory. Smalley writes the Lagrangian for Rastall theory and derived  the field equations via variational principle \cite{Smalley}.

 There are some models to explain for DE and DM: massive
gravity\cite{deRham:2010kj}, bimetric gravity, scalar-tensor gravity\cite{Ref03} and modified gravity\cite{Ref04}-\cite{Ref06}. Massive gravity is introduced by  Fierz-Pauli \cite{Fierz:1939ix} in flat space-time background and C.~de Rham and et al extend it curved spacetime\cite{deRham:2010kj}. In this paper, we
consider massive gravity in Rastall Theory and introduce the black
brane solution. This model worthwhile to study as modified of Einstein-Hilbert gravity for the unknown part of the universe.\\   

The physical description of quark-gluon plasma (QGP) is the great challenges in recent years. There are some results that QGP is strongly coupled system. Perturbation theory is not a useful way for strongly coupled theories. At the first glance Lattice QCD is usable way whereas it has some limitation. Gauge/Gravity duality \cite{Ref1}-\cite{Ref4} opens a window for solving these theories. There is a dictionary for these two different theories and we can translate the information of strongly coupled gauge theory into a weakly gravity theory and vice versa. In the long wavelength regime this duality leads to fluid/gravity duality \cite{Ref5}-\cite{Ref18}. Any fluid is characterized by some transport coefficients. One of these transport coefficients is the shear viscosity.

There are three ways to calculate shear viscosity: Green-Kubo formula, pole method and membrane paradigm. For using Green-Kubo formula we should perturb the metric and the linear response theory gives transport coefficients like shear viscosity, DC/AC conductivity and so on.\\
The final way is membrane paradigm that it means the information of fluid is located on the horizon of black brane.\\
The ratio of shear viscosity to entropy density is proportional to the inverse square coupling  of quantum thermal gauge theory. It means the stronger the coupling, the weaker the shear viscosity per entropy density.
%--------------------------------------------------------------------------
\section{Black Brane Solution in Rastall AdS Massive Gravity}
\label{sec2}

\indent The action of this theory is as follows,
\begin{equation}\label{Action1}
S=\int d^{4}  x\sqrt{-g} \Big(\frac{1}{2 \kappa'}R\,\,e^{2\sqrt{-g}\lambda' \kappa'}-2\Lambda\Big)+m^2\int d^{4}  x\sqrt{-g}\sum_{i=1}^4\,\,{c_{i}\,\,\mathcal{U}_i(g,f)},
\end{equation}

Where  $\lambda'$ and $\kappa'$ are two constant parameters. Since Lagrangian should be a scalar, these parameters recall as:
$\kappa' e^{-2\lambda' \kappa’'\sqrt{-g}} \to  \kappa$ , $\sqrt{-g} \lambda' e^{-2\lambda' \kappa' \sqrt{-g}} \to \lambda$. The parameters $\lambda$ and $\kappa$ are covariantly constant \cite{L. Smalley}. The first term is Rastall term and the last term is the massive term in (\ref{Action1}).\\
The Rastall field equation in massive gravity is as follows,
\begin{equation}\label{Action}
R_{\mu\nu}-\frac{1}{2}(R-2\Lambda)g_{\mu\nu}+{k} \lambda g_{\mu\nu}R-m^2\mathcal{\chi}_{\mu\nu}=kT_{\mu \nu},
\end{equation}
where $R$ is the Ricci scalar, $k$ the Rastall gravitational coupling constant, $\lambda$  the Rastall parameter, $\Lambda$ the cosmological constant, $G_{\mu \nu }=R_{\mu \nu}-\frac{1}{2}Rg_{\mu \nu}$ the Einstein tensor and $H_{\mu \nu }=G_{\mu \nu }+{k} \lambda g_{\mu\nu} R$  the Rastall tensor. By substituting in field equation (\ref{Action}) we have 
 \begin{equation}
H_{\mu \nu }+\Lambda g_{\mu\nu}-m^2\mathcal{\chi}_{\mu\nu}=kT_{\mu \nu}
\end{equation}
$ \mathcal{\chi}_{\mu \nu}$  is the massive term.
\begin{align}
 \mathcal{\chi}_{\mu \nu} =\frac{c_1}{2}\bigg(\mathcal{U}_1 g_{\mu \nu }-\mathcal{K}_{\mu \nu}\bigg)+\frac{c_2}{2}\bigg(\mathcal{U}_2 g_{\mu \nu }-2\mathcal{U}_1\mathcal{K}_{\mu \nu}+2\mathcal{K}^2_{\mu \nu}\bigg)+\frac{c_3}{2}\bigg(\mathcal{U}_3 g_{\mu \nu }-3\mathcal{U}_2 \mathcal{K}_{\mu \nu}\nonumber\\+6\mathcal{U}_1 \mathcal{K}^2_{\mu \nu}-6\mathcal{K}^3_{\mu \nu}\bigg)+\frac{c_4}{2}\bigg(\mathcal{U}_4g_{\mu \nu }-4\mathcal{U}_3 \mathcal{K}_{\mu \nu}+12\mathcal{U}_2 \mathcal{K}^2_{\mu \nu}-24\mathcal{U}_1 \mathcal{K}^3_{\mu \nu}+24\mathcal{K}^4_{\mu \nu}\bigg)
 \end{align}\\  
This field equations reduce to GR field equations in the limit of $\lambda \to 0$  and $k=8\pi G_N$  where $G_N$  is the Newton gravitational coupling constant.\\
In order to obtain black brane solutions, we consider the general flat symmetric spacetime metric as follows
\begin{equation}\label{Metric} 
ds^{2} =-f(r)dt^{2} +\frac{dr^{2}}{f(r)} +\frac{r^2}{l^2}(dx^2+dy^2),
\end{equation}
\\The components of Rastall tensor are as follows:
\begin{align} \label{eq04}
H_{00} &=G_{00}+{k} \lambda g_{00} R=\frac{1}{r^2}\big(f'r-1+f\big)+k\lambda R \nonumber\\
H_{11} &=G_{11}+{k} \lambda g_{11} R=\frac{1}{r^2}\big(f'r-1+f\big)+k\lambda R\nonumber\\
H_{22} &=G_{22}+{k} \lambda g_{22} R=\frac{1}{r^2}(rf'+\frac{1}{2}r^2f'')+k\lambda R\nonumber\\
H_{33} &=G_{33}+{k} \lambda g_{33} R=\frac{1}{r^2}(rf'+\frac{1}{2}r^2f'')+k\lambda R
\end{align}
where the Ricci scalar reads as
 \begin{equation}
R=\frac{1}{r^2}\big(f''r^2+4rf'-2+2f\big).
\end{equation}
The equalities $H_{00}=H_{11}$  and $H_{22}=H_{33}$  mean the symmetry properties of the Rastall tensor $H_{\mu \nu}$. \\
In (\ref{Action}), $ c_i $'s are constants and $ \mathcal{U}_i $ are symmetric polynomials of the eigenvalues of the $ 4\times4 $ matrix $ \mathcal{K}^{\mu}_{\nu}=\sqrt{g^{\mu \alpha}f_{\alpha \nu}} $
\begin{align}\label{7} 
  & \mathcal{U}_1=[\mathcal{K}]\nonumber\\
  & \mathcal{U}_2=[\mathcal{K}]^2-[\mathcal{K}^2]\nonumber\\
  &\mathcal{U}_3=[\mathcal{K}]^3-3[\mathcal{K}][\mathcal{K}^2]+2[\mathcal{K}^3]\nonumber\\
  & \mathcal{U}_4=[\mathcal{K}]^4-6[\mathcal{K}^2][\mathcal{K}]^2+8[\mathcal{K}^3][\mathcal{K}]+3[\mathcal{K}^2]^2-6[\mathcal{K}^4] 
   \end{align}

A generalized version of $ f_{\mu \nu} $ was proposed in {\cite{Cai:2014znn} with the  form
$ f_{\mu \nu} = \frac{c_0^2}{l^2}diag(0,0,1,1)$.

The values of $ \mathcal{U}_i $ are calculated as below,
\begin{align}\label{9} 
  & \mathcal{U}_1=\frac{2c_{0}}{r}, \,\,\,  \,\,\, \mathcal{U}_2=\frac{2c_0^2}{r^2},\,\,\,\,\mathcal{U}_3=0,\,\,\,\,\mathcal{U}_4=0\nonumber.
   \end{align}

$f(r)$ is found by inserting the  ansatz (\ref{Metric}) in field equation Eq.(\ref{Action}),
\begin{equation}
 f(r)=1-\frac{b}{r}-\frac{\Lambda}{3} r^2+m^2l^2 \bigg(\frac{c_0c_1}{2}r+c_0^2c_2\bigg).
\end{equation}
 Event horizon is where $f(r_+)=0$ and we can find $b$  by applying this condition,
  \begin{equation}
b= r_+\Bigg[1-\frac{\Lambda}{3} r_+^2+m^2l^2\bigg(\frac{c_0c_1}{2}r_++c_0^2c_2\bigg)\Bigg]
 \equiv r_+\bigg(1-\frac{\Lambda}{3} r_+^2+\Delta\bigg)\\
\end{equation}
where $\Delta$ is,
 \begin{equation}
\Delta \equiv m^2l^2\bigg(\frac{c_0c_1}{2}r_++c_0^2c_2\bigg)
\end{equation}
By substituting $b$ in $f(r)$ we have,
 \begin{equation}\label{f}
f(r)=\frac{1}{r}\bigg[(r-r_+)-\frac{ \Lambda}{3}(r^3-r_+^3)+m^2l^2c_0^2c_2(r-r_+)+m^2l^2\frac{c_0c_1}{2}(r^2-r_+^2)\bigg] .
\end{equation}
Cosmological constant depends to the Rastall parameter as $\Lambda=\frac{1}{3}\frac{{\rho}_0}{4\lambda-1}$. The solution is the same as \cite{Cai:2014znn}  when $\lambda=0$  and $\rho_{0}=\frac{9}{l^2}$.\\
 Hawking temperature is defined by,
\begin{equation}
T=\frac{f '(r_{+} )}{4\pi }=\frac{1}{4\pi }\bigg[\frac{1}{ r_+}-\Lambda r_++\frac{m^2l^2c_0^2 c_2}{r_+}+m^2l^2c_0 c_1\bigg].
\end{equation}
The entropy can be found by using Hawking-Bekenstein formula,
\begin{eqnarray}
%\begin{align*}
A&=&\int d^{2} x \sqrt{-g} |_{r=r_+,t=cte}= \frac{r_{+}^{2} V_{2}}{l^{2}} \nonumber\\
S&=&\frac{A}{4G} =\frac{r_{+}^{2} V_{2} }{4l^{2}G} \nonumber\\
s&=&\frac{S}{V_{2} } =\frac{4\pi r_{+}^{2} }{l^{2}}
%\end{align*}
\end{eqnarray}
where $V_{2}$ is the volume of the constant $t$ and $r$ hyper-surface with radius $r_{+}$ and in the last line we used $\frac{1}{16\pi G} =1$ so $\frac{1}{4G} =4\pi$.\\
\section{Shear Viscosity to Entropy Density}
 \label{sec4}
The black brane solution is,
 \begin{align}\label{met2}
   ds^{2} =-f(r)dt^{2} +\frac{dr^{2} }{f(r)} +\frac{r^2}{l^2}(dx^2+dy^2)
  \end{align} 
 Where $f(r)$ given in Eq.\ref{f}  and $r$ is the radial coordinate that put us from bulk to boundary.\\

  Now we calculate shear viscosity to entropy density via membrane paradigm method. Consider the metric as follow
\begin{equation}
ds^{2}=g_{00}(r)dt^{2} +g_{rr}(r)dr^{2}+g_{xx}(r)\sum_{i=1}^{p}(dx^{i})^{2}
\end{equation}
Since effective hydrodynamics in field theories is constructed in terms of conserved currents and energy-momentum tensor of the theory we consider the small fluctuations of the black brane background as $g_{\mu \nu} \to g_{\mu \nu}+ h_{xy}$ for calculating of shear viscosity. Where $h_{xy}$ is a small perturbation.\\
By following the procedure of  \cite{Ref19}  the formula for $\frac{\eta}{s} $ is as,
\begin{equation}\label{eq43}
\frac{\eta }{s} =T\frac{\sqrt{-g(r_{+})}}{\sqrt{-g_{00}(r_{+})g_{rr}(r_{+})}}\int_{r_{+}}^{\infty}{\frac{-g_{00}(r)g_{rr}(r)}{g_{xx}(r)\sqrt{-g(r)}}} dr.
\end{equation}
By considering the metric Eq.(\ref{met2}) and applying Eq.(\ref{eq43}) we have,
 \begin{equation}
 \frac{\eta }{s} =Tr_{+}^{2}\int_{r_{+}}^{\infty}\frac{dr}{r^{4}}=\frac{T}{3r_{+}}=\frac{1}{12\pi r_+}\bigg[\frac{1}{ r_+}-\Lambda r_++\frac{m^2l^2c_0^2 c_2}{r_+}+m^2l^2c_0 c_1\bigg]
 \end{equation}
Since hydrodynamics is an effective describtion of field theory in long-wavelength limit we should consider large $ r_{+} $, so by applying this criterion,
  \begin{equation}\label{mem-para}
   \frac{\eta }{s}=-\frac{\Lambda}{12\pi}
   \end{equation}
By Substituting $\Lambda=\frac{1}{3}\frac{{\rho}_0}{4\lambda-1}$ in the valu of   $ \frac{\eta }{s}$ is as 
\begin{equation}\label{mem-para}
   \frac{\eta }{s}=-\frac{\Lambda}{12\pi}=\frac{1}{4\pi}\frac{{\rho}_0}{9-36\lambda}
   \end{equation}
The result depends on the the Rastall parameter $\lambda$ and vaccum energy density ${\rho}_0$.  If ${\rho}_0 < 9-36\lambda$ or ($\Lambda > -3 $) then $ \frac{\eta }{s}$ is less than $\frac{1}{4\pi }$  and if ${\rho}_0 = 9-36\lambda$ or ($\Lambda=- 3 $) then $ \frac{\eta }{s} = \frac{1}{4\pi }$.\\ 
 The spacetime is AdS so the cosmological constant should be negative. It means that $ \frac{\eta }{s} \leq \frac{1}{4\pi }$ for this model.

%--------------------------------------------------------------------------
% \section{DC and AC conductivities}

%\large{For calculating these}

%--------------------------------------------------------------------------
 \section{Conclusion}

\noindent   We have used the fluid/gravity duality to obtain information about the dual Rastall gravity in 4 dimensions. We showed the KSS bound depends on the vacuum energy density ${\rho}_0$ . KSS bound saturates for Einstein-Hilbert gravity \cite{Ref16}-\cite{Ref18} but this bound violates for higher derivative gravities like the Gauss-Bonnet gravity \cite{Ref23}-\cite{Parvizi:2017boc}. Our result shows that this theory behaves effectively like higher derivative gravities.

%--------------------------------------------------------------------------
\vspace{1cm}
\noindent {\large {\bf Acknowledgment} }    The author would like to thank the referee of MPLA for the valuable comments which helped to improve the manuscript.

%--------------------------------------------------------------------------

\end{document}